\documentclass[twocolumn,preprintnumbers,amsmath,amssymb,plb]{revtex4}

\usepackage{epsfig}
\usepackage{graphicx}
\usepackage{graphics}
\usepackage{dcolumn}
\usepackage{bm}

\begin{document}

\title{\Large\bf\boldmath The Measurements of $J/\psi \rightarrow p \bar{p}$ }
\author{J.~Z.~Bai$^{1}$,        Y.~Ban$^{10}$,         J.~G.~Bian$^{1}$,
X.~Cai$^{1}$,         J.~F.~Chang$^{1}$,       H.~F.~Chen$^{16}$,    
H.~S.~Chen$^{1}$,       H.~X.~Chen$^{1}$,      J.~Chen$^{1}$,        
J.~C.~Chen$^{1}$
% ...
\\(BES Collaboration)\\ 
$^1$ Institute of High Energy Physics, Beijing 100039, People's Republic of
     China\\
$^2$ China Center of Advanced Science and Technology, Beijing 100080,
     People's Republic of China\\
$^3$ Guangxi Normal University, Guilin 541004, People's Republic of China\\
$^4$ Henan Normal University, Xinxiang 453002, People's Republic of China\\
$^5$ Huazhong Normal University, Wuhan 430079, People's Republic of China\\
$^6$ Hunan University, Changsha 410082, People's Republic of China\\                                                  
$^7$ Liaoning University, Shenyang 110036, People's Republic of China\\
$^{8}$ Nanjing Normal University, Nanjing 210097, People's Republic of China\\
$^{9}$ Nankai University, Tianjin 300071, People's Republic of China\\
$^{10}$ Peking University, Beijing 100871, People's Republic of China\\
$^{11}$ Shandong University, Jinan 250100, People's Republic of China\\
$^{12}$ Shanghai Jiaotong University, Shanghai 200030, 
        People's Republic of China\\
$^{13}$ Sichuan University, Chengdu 610064,
        People's Republic of China\\                                    
$^{14}$ Tsinghua University, Beijing 100084, 
        People's Republic of China\\
$^{15}$ University of Hawaii, Honolulu, Hawaii 96822\\
$^{16}$ University of Science and Technology of China, Hefei 230026,
        People's Republic of China\\
$^{17}$ Wuhan University, Wuhan 430072, People's Republic of China\\
$^{18}$ Zhejiang University, Hangzhou 310028, People's Republic of China\\
$^{\ast}$ Visiting professor to University of Michigan, Ann Arbor, MI 48109 USA 
}

\noindent
\begin{abstract}
The process $J/\psi \rightarrow p \bar{p}$ is studied using 57.7$\times 10^6$
$J/\psi$ events collected with the BESII detector at the Beijing Electron 
Positron Collider(BEPC). The branching ratio is determined to 
be $Br(J/\psi \rightarrow p \bar{p})=(2.26 \pm 0.01 \pm 0.14)\times 10^{-3}$,
and the angular distribution is well described by
$\frac{dN}{d\cos\theta_p}=1+\alpha\cos^2\theta_p$ with $\alpha=0.676 \pm
0.036 \pm 0.042$,
where $\theta_p$ is the angle between proton and beam direction.
The obtained $\alpha$ value is in good agreement 
with predictions of first-order QCD.

\end{abstract}
 
\maketitle

\clearpage
\section{Introduction}
\label{secintro}
The $J/\psi$ meson is interpreted as a bound state of a charmed quark and 
charmed antiquark ($c\bar{c}$). The decay process $J/\psi \rightarrow p \bar{p}$ 
is an octet-baryon-pair decay mode that has been measured by DM2~\cite{dm2}, 
DASP~\cite{dasp}, Mark I~\cite{mark1}, Mark II~\cite{mark2} and Mark III~\cite{mark3}. 

In general, the angular distribution 
of $J/\psi \rightarrow B \bar B$ can be written as
\[
\frac{dN}{d\cos\theta_B} \propto 1+\alpha\cos^2\theta_B,
\]
where $\theta_B$ is the angle between the direction of the 
omitting baryon and the $e^ -e^+$ beam direction.
Different theoretical models based on 
first-order QCD give different predictions for the value of
$\alpha$~\cite{theo1,theo2,theo3}. Brodsky and Lepage~\cite{theo1} 
assumed that the reaction is a process in the asymptotic region where 
all quark and baryon masses can be neglected. 
Claudson, Glashow, and Wise~\cite{theo2} included the baryon masses 
in their calculation, which significantly changed the previous prediction. 
Carimalo~\cite{theo3} considered the effect of non-zero quark masses. 
The branching ratio and corresponding angular distribution 
will provide some insight into the details of the
baryon structure~\cite{theo1,theo2,theo3,theo4,theo5}. 
Chernyak and Zhitnitsky~\cite{theo4} have investigated
QCD sum rules for nucleon wave functions and calculated the branching
ratio of the discussed decay in terms of the wave function.
Ping, Chiang and Zou~\cite{theo5} studied the branching ratio as well as the
angular distribution for $J/\psi \rightarrow B \bar{B}$ decay processes
in naive quark model.
Table~\ref{T1} shows the measured branching ratio and $\alpha$ value 
of $J/\psi \rightarrow p\bar{p}$ by previous experiments. 
Table~\ref{T2} shows the different theoretical 
predictions on $\alpha$ value.

\begin{table}[htbp]
\begin{center}
      \caption{Previous experimental results for  $J/\psi \rightarrow p\bar{p}$}
      \label{T1}
      \begin{tabular}{|c|c|c|c|}\hline\hline
      \multicolumn{4}{|c|}{measured value about Br. and $\alpha$}\\\hline
       Coll. & $ N^{obs}$ &\multicolumn{1}{c|}{ Br$(\times10^{-3})$}
                          &\multicolumn{1}{c|}{ $\alpha$}\\\hline
       Mark1 & 331        &\multicolumn{1}{c|}{ 2.2$\pm$0.2}
                          &\multicolumn{1}{c|}{1.45$\pm$0.56}\\\hline
       Mark2 & 1420       &\multicolumn{1}{c|}{ 2.16$\pm$0.07$\pm$0.15}
                          &\multicolumn{1}{c|}{ 0.61$\pm$0.23}\\\hline
       Mark3 &            &\multicolumn{1}{c|}{ 1.91$\pm$0.03$\pm$0.16}
                          &\multicolumn{1}{c|}{ 0.58$\pm$0.14}\\\hline
       DASP  & 133        &\multicolumn{1}{c|}{2.5$\pm$0.4}
                          &\multicolumn{1}{c|}{ 1.70$\pm$1.70}\\\hline
       DM2   &            &\multicolumn{1}{c|}{ 1.91$\pm$0.04$\pm$0.30}
                          &\multicolumn{1}{c|}{ 0.62$\pm$0.11}\\\hline
       PDG\cite{pdg2002}  &            &\multicolumn{1}{c|}{ 2.12$\pm$0.10}  
                          &\multicolumn{1}{c|}{0.63$\pm$0.08(avg.)}\\\hline\hline
\end{tabular}
\end{center}
\end{table}

\begin{table}[htbp]
\begin{center}
      \caption{Different theoretical predictions on $\alpha$ value}
      \label{T2}
      \begin{tabular}{|c|c|}\hline\hline
      \multicolumn{2}{|c|}{Theoretical values of $\alpha$}\\\hline
      $\alpha$             &authors                       \\\hline   
      1.0\cite{theo1}      &Brodsky and Lepage            \\\hline
      0.46\cite{theo2}     &Claudson, Glashow and Wise    \\\hline
      0.69,0.70\cite{theo3}&Carimalo                      \\\hline
\end{tabular}
\end{center}
\end{table}

In this paper, we report high precision measurements 
of the branching ratio of $J/\psi \rightarrow p \bar{p}$ and 
corresponding $\alpha$ value. The measurements are based on an
analysis of 57.7$\times 10^6 J/\psi$~\cite{n-jpsi}events registered
in the BESII detector at BEPC.

BESII is a large solid-angle magnetic spectrometer that is described  
in Ref.~\cite{bes2}. Charged particle momenta are determined with a 
resolution of $\sigma_p/p$=$1.78\%$ $\sqrt{1+p^2(\rm{GeV}^2)}$ in a 40-layer 
cylindrical main drift chamber(MDC). Particle identification is accomplished by 
specific ionization ($dE/dX$) measurements in the drift chamber and 
time-of-flight (TOF) measurements in a barrel-like array of 48 
scintillation counters. The $dE/dX$ resolution is $\sigma_{dE/dX}=8.0\%$;
the TOF resolution for Bhabha events  is $\sigma_{TOF}=180$ ps. These
two systems 
independently provide more than $3\sigma$ separation of protons from any 
other charged  particle species for the momentum range relevant
to $J/\psi \rightarrow p \bar{p}$ decays. Radially outside of the 
TOF counters is a 
12-radiation-length
barrel shower counter (BSC) comprised of gas proportional tubes 
interleaved with lead sheets. The BSC measures the energies and directions
of photons with resolution of $\sigma_E/E \simeq 21\%/\sqrt{E(\rm{GeV})}$,
$\sigma_{\phi}=7.9$ mrad, and $\sigma_{z}=2.3$ cm. The iron flux 
return of the magnet is instrumented with three double layers of counters 
that are used to identify muon.

In this analysis, a GEANT3 based Monte Carlo package(SIMBES) with 
detailed consideration of the detector performance (such as dead 
electronics channels) is used. The consistency between data and Monte Carlo
has been carefully checked in many high purity physics channels, and the 
agreement is reasonable.

\section{Event Selection}
\label{secsele}
Events are selected with two and only two well 
reconstructed and oppositely charged tracks.
The polar angle of the track, $\theta$, 
is required to be in the range $|\cos\theta| < 0.8$. 
In order to remove the cosmic rays, the difference of the flying time  
between the positive and negative track, $|t_{+}-t_{-}|$, is required to 
be less than 4.0 ns. Information from the TOF system is used to  
identify protons. Both tracks are required to be
not identified as a $\pi$ or a $K$ by the TOF, $\it i.e.$
the measured flying time and the expected flying time assuming the track 
to be proton, pion, or kaon should satisfy the following requirement, $|t_{meas}-t_{exp}(p)|<|t_{meas}-t_{exp}(\pi)|$ and 
$|t_{meas}-t_{exp}(p)|<|t_{meas}-t_{exp}(K)|$.
Meanwhile, since $J/\psi \rightarrow p \bar{p}$ is a
two-body decay, the $p\bar{p}$ pairs are back-to-back; 
we require the two-track opening angle to be larger than 
$175^{\circ}$. Finally, for both tracks, the absolute difference between the 
measured momentum and its expected value, 1.232 GeV/c, 
is required to be less than 
110 MeV/c.

\begin{figure*}[htbp]
\begin{center}
\includegraphics[width=15.0cm,height=10.cm]{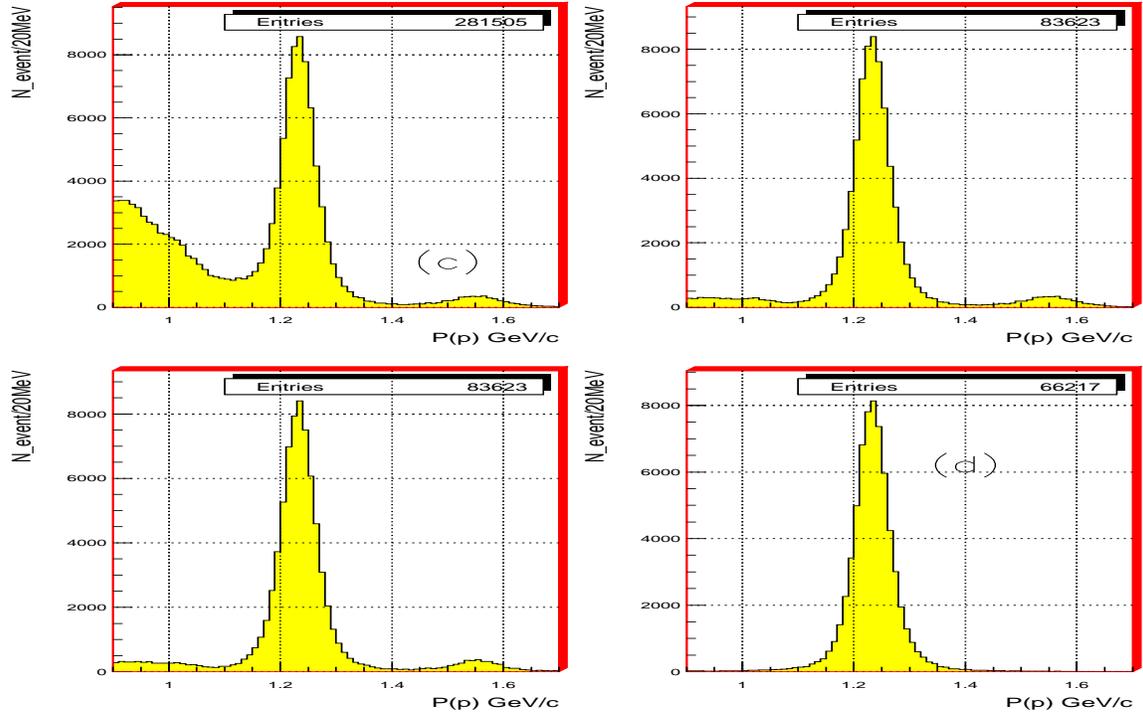}
\caption[]{The proton momentum distribution for events surviving
 different selection criteria:
           (a) after particle ID and cosmic ray veto;
           (b) after the back-to-back requirement;
           (c) the corresponding antiproton momentum distribution;
           (d) the proton momentum distribution after selecting the 
               antiproton momentum as shown in(c).}
\label{ppcorr1}
\end{center}
\end{figure*}

Figures~\ref{ppcorr1}(a)-(d) show the proton momentum 
distribution for events surviving different selection criteria: 
Fig.~\ref{ppcorr1}(a) is after the particle ID and cosmic rays veto;
Fig.~\ref{ppcorr1}(b) is after the back-to-back requirement;
Fig.~\ref{ppcorr1}(c) is the antiproton momentum after
the same cuts; and Fig.~\ref{ppcorr1}(d) is 
the proton momentum distribution after selecting the momentum 
of antiproton with $\pm$110 MeV/c cut. From this last figure it is 
apparent that most of the background has been rejected. 

\section{Background estimation}
\label{secback}
Two methods are used  to estimate the background level.
Method one uses a $J/\psi \rightarrow anything$ MC sample that is produced 
using the Lund-charm~\cite{lund-charm} generator. From
this sample the level of background events in the data 
is determined to be about 
$1.5\%$ of all selected events. This background is mainly 
due to $J/\psi \rightarrow \gamma p \bar{p}$, 
$J/\psi \rightarrow \pi^0 p \bar{p}$ and 
$J/\psi \rightarrow \gamma \eta_c,~\eta_c \rightarrow p \bar{p}$. 
Before application of the proton momentum requirement, the background 
level is about 1.7$\%$.
This study indicates that 
there is little background for high momentum values, but some at 
lower momenta, 
as can be seen in Fig.~\ref{ppcorr1} (d).
 
 In method two, possible two-body background channels, such as 
$J/\psi \rightarrow e^+ e^-$, $J/\psi \rightarrow \mu^+ \mu^-$, 
$ J/\psi \rightarrow K^+ K^-$, $J/\psi \rightarrow \pi^+ \pi^-$, 
$J/\psi \rightarrow \gamma p \bar{p}$, $J/\psi \rightarrow \pi^0 p\bar{p}$, 
and $J/\psi \rightarrow \gamma \eta_c$, $\eta_c \rightarrow p \bar{p}$ 
are considered and generated according to the branching ratios 
listed in PDG(2002)~\cite{pdg2002}. The total background level for 
these events surviving the $J/\psi \rightarrow p \bar p$ event 
selection criteria is about $0.7\%$. 
 
Since the background level is very small, we do not subtract 
the background but use $1.5\%$ as the uncertainty of the background 
in the branching ratio measurement.

\section{Efficiency correction}
\label{seccorr}
Since there are imperfections in Monte Carlo simulation, particularly
when it simulates the tail of momentum resolution and TOF resolution, 
which will affect the efficiencies of momentum cuts and particle ID.
In order to reduce systematic error
as much as possible, the correction on MC efficiency is needed. 
Since both momentum resolution and TOF resolution are $\cos\theta$ 
dependent, when we put cuts on momentum and TOF, the efficiency
will depend on the angle $\theta$.
A reweighting method is used to do the efficiency correction. 
We define a $\cos\theta$ dependent weight factor (or correction factor) 
$wt_j(\cos\theta)$ as:
\[
wt_j(\cos\theta)=\epsilon^{data}_j(\cos\theta)/\epsilon^{mc}_j(\cos\theta),
\]
where $j$ represents the $j$th selection requirement, $\theta$ is the 
polar angle of $p$ or $\bar p$, and $\epsilon$ stands for efficiency. 
When the weight functions are obtained, phase space MC generator is used and 
500,000 events are generated. Since the weight is obtained
in each small bin of $\cos\theta$, the weight function will not depend
very much on the actual angular distribution of generated events.
To determine the weight for a given variable, the other, 
uncorrelated requirements are made more stringent to provide a pure 
sample.  When the correlation among different selection criteria is so 
small that it can be ignored, the efficiency of the data can be expressed 
by the reweighted MC efficiency as:
 \begin{eqnarray*}
 \epsilon_{data}(\cos\theta)
 &=&\epsilon_X\prod^n_{j=1}(\epsilon^{data}_j(\cos\theta)) \nonumber \\
 &=& \epsilon_X\prod^n_{j=1}(\epsilon^{mc}_j(\cos\theta)\times 
    wt_j(\cos\theta)) \nonumber \\
 &=&\epsilon_X\prod^n_{j=1}\epsilon^{mc}_j(\cos\theta)\prod^n_{j=1}
    wt_j(\cos\theta) \nonumber \\
 &=& \epsilon_{mc}(\cos\theta)\times wt_{tot}(\cos\theta),
 \label{eq:0402}
 \end{eqnarray*}
where $\epsilon_X$ is the part of the efficiency which is not corrected;
it is due to  track reconstruction, geometry acceptance, etc, and its
uncertainty can be determined using another method(see Sec.~\ref{secsyst} 
below). The total correction factor $wt_{tot}$ is:
\begin{eqnarray*}
 & & wt_{tot}(\cos\theta) = \nonumber \\
 & & wt_p(\cos\theta)\times wt_{\bar{p}}(\cos\theta)
\times wt_{pid}(\cos\theta) \times wt_{\bar{p}id}(\cos\theta),
\end{eqnarray*}
where $wt_{p}$ and $wt_{\bar{p}}$ are the weights for the proton and antiproton
momentum requirements, while $wt_{pid}$ and $wt_{\bar{p}id}$ are the weights
for the particle identification requirements on proton and antiproton.

\section{Angular distribution}
\label{secangu}
With $\epsilon_{mc}(\cos\theta)$ denoting to the efficiency 
obtained from Monte Carlo and $wt_{tot}(\cos\theta)$ for the total 
correction function of the efficiency, we fit the angular distribution 
of the data with the function $f(\cos\theta)$, 
\[
 f(\cos\theta)=A\times(1+\alpha \cos^2\theta)\times 
               \epsilon_{mc}(\cos\theta)\times 
               wt_{tot}(\cos\theta)
\]
where $A$ is a constant.

\begin{figure*}[htb]
\centering
\includegraphics[width=14.0cm]{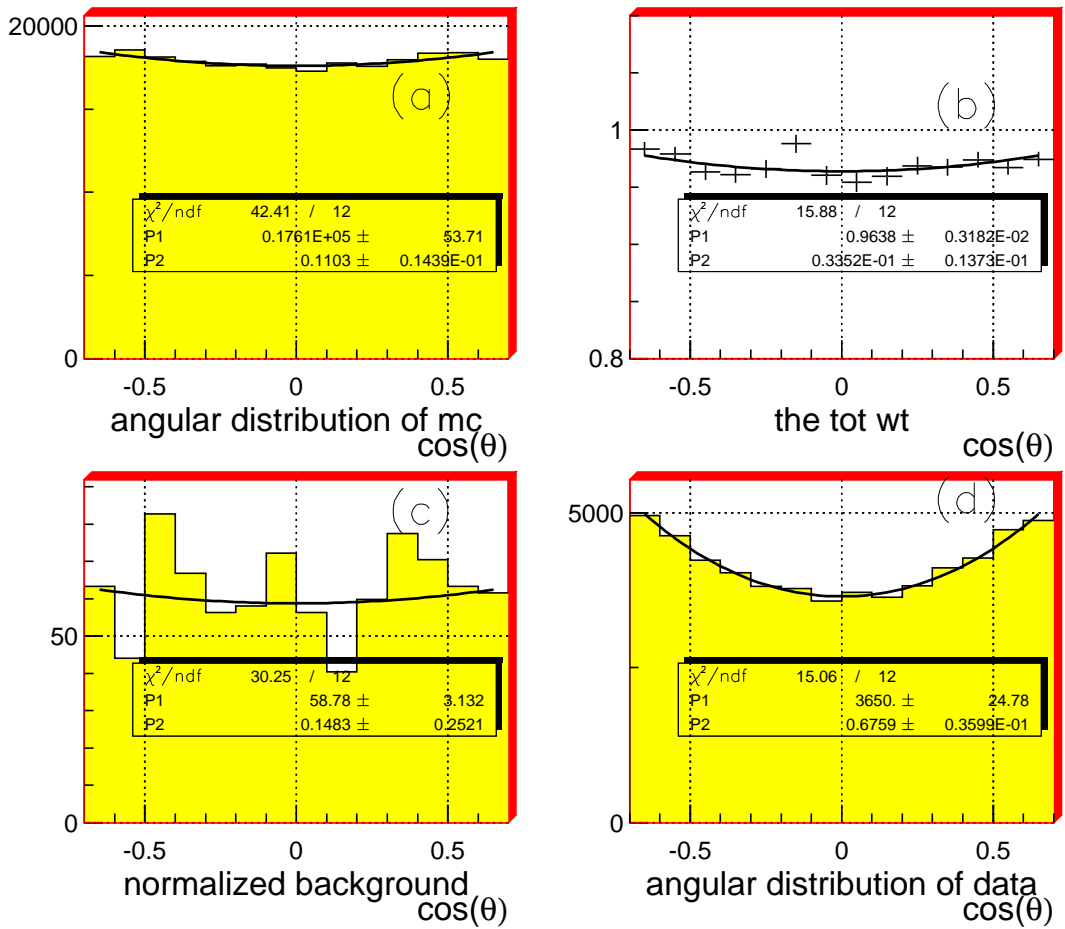}
\caption{The (a) angular distribution for MC events; 
         (b) total weight curve; 
         (c) angular distribution of the background; and 
         (d) angular distribution of the data.}
\label{angular}
\end{figure*}

Fig.~\ref{angular}(a) shows the angular distributions for 
500,000 phase space MC events and Fig.~\ref{angular}(c) for 
the sideband background which is from $p$,$\bar{p}$ momentum distribution 
from 1.00 GeV/c to 1.10 GeV/c. 
Fig.~\ref{angular}(b) shows the total weight curve and 
Fig.~\ref{angular}(d) shows the result of 
fitting to the angular distribution of the data.
The fit to the angular distribution of data gives 
\[
 \alpha = 0.676 \pm 0.036,
\]
where the error is statistical only.

\section{Branching ratio of $J/\psi \rightarrow p \bar{p}$}
\label{secbran}
 After the final selection, the number of signal events is 
 \[
 n^{obs}_{data}=63316 
 \]
 The Monte Carlo generator, which includes the angular distribution of
$\frac{dN}{d\cos\theta_p}=1+\alpha\cos^2\theta_p$ with $\alpha = 0.68$
obtained from above fit, is used and 50,000 events are generated 
to estimate the selection efficiency for branching ratio calculation. 
The MC-determined selection efficiency $\epsilon_{mc}$, including the 
effect of reweighting, is $\epsilon_{mc}=(48.53 \pm 0.31)\%$ and the
branching ratio is determined to be:
 \[
 Br(J/\psi \rightarrow p \bar{p})= \frac{n^{obs}_{data}}
 {N^{tot}_{J/\psi}\times\epsilon_{mc}}=(2.26 \pm 0.01 ) 
 \times 10^{-3}, 
 \]
where $N^{tot}_{J/\psi}$ is the total number of $J/\psi$ 
in the data sample ($57.7 \times 10^6$)~\cite{n-jpsi}. 
The error is statistical only.

 \section{Systematic error}
\label{secsyst}
 \subsection{Systematic error of angular distribution}
 
When the fit parameters of the weight curve are changed 
by $1\sigma$, the fitted value for $\alpha$ changes by 2.6$\%$. 
This is taken as a systematic error. Another systematic
error from tracking reconstrucion is determined by using different 
MDC wire resolution models in the 
MC simulations. This affects the $\alpha$ value by 5.2$\%$. 
From a $p$,$\bar{p}$ momentum sideband study, 
the uncertainty from background, including background level and background shape, 
is estimated to be 2.2$\%$. The uncertainties from other sources 
such as from the fitting function of $\epsilon_{mc}(\cos\theta)$ are negligible. 
Adding these contributions in quadrature
gives a  total effect about 6.2$\%$, i.e. 
 \[
 \alpha = 0.676 \pm 0.036 \pm 0.042, 
 \] 
 where the first error is statistical and the second systematic.

\subsection{Systematic error of the branching ratio}
Systematic errors on the branching ratio measurement mainly come 
from MC statistics, the efficiency uncertainties of the particle ID, 
the momentum selection, the MDC wire resolution,
the uncertainty in $\alpha$, uncertainties in the background level, 
and the total number of $J/\psi$ events. 
 
The MC statistics gives a systematic error about $0.6\%$. 
As discussed in Sec.~\ref{secback}, we do not subtract background but
use $1.5\%$ as the uncertainty from the background in the branching ratio 
measurement. Reweighting is used to correct the PID and momentum cut efficiencies. 
When the weight curve is changed by $1\sigma$, 
the branching ratio changes by $0.3\%$.
Different wire resolution models used in the MC produce shifts of $3.6\%$ 
in the branching ratio.
When the channel $J/\psi \rightarrow p \bar{p}$ is simulated, the $\alpha$
value is an input parameter. If we change the value by $1\sigma$, the
efficiency changes by $0.6\%$.
The total number of $J/\psi$ events is determined to be 
$(57.70\pm 2.72)\times  10^{6}$~\cite{n-jpsi}; we use $4.7\%$ as the 
systematic error from this source.
 
Combining the uncertainties from all sources in quadrature, 
the total effect on the branching ratio of $J/\psi \rightarrow p \bar{p}$ 
is about $6.2\%$, i.e. 
 \[
   Br(J/\psi \rightarrow p \bar{p}) = (2.26 \pm 0.01 \pm 0.14)\times 10^{-3}, 
 \]
 where the first error is statistical and the second systematic. 

 Table~\ref{T3} lists the systematic errors from all sources. 
\begin{table}[htpb]
\begin{center}
    \renewcommand{\arraystretch}{1.5}
    \caption {The systematic errors }
    \label{T3}
    \begin{tabular}{|c|c|c|}\hline\hline
    Sources           & Effect on $\alpha$ & Effect on Br.\\\hline \hline
    MC statistics     & --              & 0.6$\%$ \\\hline 
    background        & 2.2$\%$         & 1.5$\%$ \\\hline
    wire resolution   & 5.2$\%$         & 3.6$\%$ \\\hline
    reweighting       & 2.6$\%$         & 0.3$\%$ \\\hline
    $\alpha$ value    & --              & 0.6$\%$ \\\hline
    $J/\psi$ number   & --              & 4.7$\%$ \\\hline
    total effect      & 6.2$\%$         & 6.2$\%$ \\\hline
    \end{tabular}
\end{center}
\end{table}

\section{Summary }
\label{secsumm}
 The channel $J/\psi \rightarrow p \bar{p}$ is studied with $57.7 \times 10^6$
 $J/\psi$~\cite{n-jpsi} events. The branching ratio and angular  
distribution are measured. The branching ratio is within one 
$\sigma$ of the PDG world average value and the angular distribution 
$\alpha$ is in good agreement with the theoretical prediction 
that includes the effects of non-zero quark masses~\cite{theo3}.
 
\section{Acknowledgment}
\label{secackn}
 We acknowledge the strong efforts of staff of the BEPC and the helpful 
 assistance from the members of the IHEP computing center. 
 This work is supported in part by the National Natural Science Foundation
of China under contracts Nos. 19991480,10225524,10225525, the Chinese Academy 
of Sciences under contract No. KJ 95T-03, the 100 Talents Program of CAS 
under Contract Nos. U-11, U-24, U-25, and the Knowledge Innovation Project of 
CAS under Contract Nos. U-602, U-34(IHEP); by the National Natural Science 
Foundation of China under Contract No.10175060(USTC); and by the Department 
of Energy under Contract No.DE-FG03-94ER40833 (U Hawaii).


\begin{thebibliography}{99}
 \bibitem{dm2}   D.Pallin {\it et al.}, Nucl. Phys. {\bf B292}, 653 (1987).
 \bibitem{dasp}  R.Brandelik {\it et al.}, Phys. {\bf C1}, 233 (1976). 
 \bibitem{mark1} I.Peruzzi {\it et al.}, Phys. Rev. {\bf D17}, 2901 (1978).
 \bibitem{mark2} M.W.Eaton {\it et al.}, Phys. Rev. {\bf D29}, 804 (1984)
 \bibitem{mark3} J.S.Brown, PhD Thesis, University of Washington UMI 84-19117-mc (unpublished). 
 \bibitem{theo1} S.J.Brodsky, G.P.Lepage, Phys. Rev. {\bf D24}, 2848 (1981).
 \bibitem{theo2} M.Claudson, S.L.Glashow, M.B.Wise, Phys. Rev. {\bf D25}, 
                 1345 (1982).
 \bibitem{theo3} C.Carimalo, Int.J.Mod. Phys. {\bf A2}, 249 (1987).
 \bibitem{theo4} V.L.Chernyak and I.T.Zhitnitsky, Nucl. Phys. {\bf B246}, 52 (1984).
 \bibitem{theo5} R.G.Ping, H.C.Chiang, B.S.Zou, Phys. Rev. {\bf D66} (2002) 054020.
 \bibitem{bes2}  J.Z.Bai {\it {et al.}} (BES Collaboration), Nucl. Instr. Meth., {\bf A458}, 627 (2001).
 \bibitem{lund-charm} J.C.Chen {\it et al.}, Phys. Rev. {\bf D62}, 034003 (2000)
 \bibitem{n-jpsi} Fang Shuangshi {\it et al.}, HEP\&NP {\bf 27}, 277 (2003) (in Chinese).
 \bibitem{pdg2002}  K. Hagiwara {\it et al.} (Particle Data Group), Phys. Rev. {\bf D66}, 010001 (2002).

 \end{thebibliography}
\end{document}